\documentclass[draft]{agujournal2019}
\usepackage{url} 
\usepackage{lineno}
\usepackage[inline]{trackchanges} 
\usepackage{soul}
\usepackage{amsmath}
\usepackage{ragged2e}
\usepackage{float}
\usepackage{graphicx}
\usepackage{multirow}
\justifying
\nolinenumbers
\draftfalse

\journalname{JGR: Planets}

\begin{document}

%
%


\title{
Dynamos driven by top-heavy double-diffusive convection in the strong-field regime}

%
%



\authors{Wei Fan\affil{1} and Yufeng Lin\affil{1,2}}



\affiliation{1}{Department of Earth and Space Sciences, Southern University of Science and Technology, Shenzhen, China}
\affiliation{2}{Center for Complex Flows and Soft Matter Research, Southern University of Science and Technology, Shenzhen,China}




\correspondingauthor{Yufeng Lin}{linyf@sustech.edu.cn}



\begin{keypoints}
\item We performed hydrodynamic and dynamo simulations of top-heavy double-diffusive convection in a spherical shell at large Lewis number. 

\item The convective flow morphology strongly depends on the nature of the buoyancy sources in hydrodynamic simulations.

\item Dynamo actions in the strong-field regime show minimal differences between the double-diffusive and co-density models.

\end{keypoints}

%
%

%
%


\begin{abstract}
The magnetic fields of terrestrial planets are generated in their liquid cores through dynamo action driven by thermal and compositional convection. The coexistence of these two buoyancy sources gives rise to double-diffusive convection (DDC) due to the contrast between thermal and compositional diffusivities. However, most dynamo simulations adopt the co-density model, where the two diffusivities are assumed to be equal.
In this study, we performed both hydrodynamic and dynamo simulations of top-heavy DDC in a rotating spherical shell with the Lewis number $Le=100$,  and compared them with corresponding co-density models. In the hydrodynamic regime, the convective flow morphology is strongly influenced by the nature of the buoyancy sources. However, our dynamo simulations in the strong-field regime demonstrate that the co-density and DDC models yield qualitatively similar magnetic fields at comparable magnetic Reynolds numbers, albeit with some differences in detail. 
These numerical models further justify the use of the co-density model in planetary dynamo simulations. Finally, we demonstrate that dynamo models based on DDC and co-density produce similar magnetic fields and secular variations at the core-mantle boundary. This suggests that it may not be possible to distinguish the buoyancy sources responsible for planetary dynamos based solely on magnetic field observations.
\end{abstract}

\section*{Plain Language Summary}
Planets like Earth generate their magnetic fields through movements in their liquid cores, a process known as dynamo action. These movements are driven by heat and compositional differences in the core, leading to a phenomenon called double-diffusive convection. While previous studies often assumed that heat and composition spread equally in the core, this study explores what happens when their spread rates differ by two orders of magnitude, which is more realistic for planetary interiors.
Using advanced simulations, we compared models that account for these differences with simpler ones where the spread rates are assumed equal. We found that while the flow patterns inside the core differ in non-magnetic conditions, the magnetic fields generated in both models are surprisingly similar, especially when strong magnetic fields are present. This suggests that the simpler models can still provide proper insights into planetary dynamos.
Importantly, the results show that we may not be able to identify the exact sources of core movements that generate planetary magnetic fields using only magnetic field data. This finding helps refine our understanding of how planetary magnetic fields are generated and supports the use of simplified models in future studies.

\section{Introduction}\label{sec:intro}
The magnetic fields of terrestrial planets are generated by convective motions in the metallic liquid core through dynamo action \cite{Tikoo2022}. Depending on the thermal evolution history of a planet, convection in the core can be driven by thermal buoyancy, compositional buoyancy, or a combination of both \cite{Landeau2022sustaining}. Thermal convection (TC) results from the superadiabatic temperature gradient, which is produced by the latent heat released during the solidification of iron at the inner core boundary (ICB) and by heat extraction from the mantle at the core-mantle boundary (CMB). Compositional convection (CC) is driven by the persistent crystallization of the inner core, which releases lighter elements such as silicon or oxygen \cite{hirose2013composition}. Prior to the formation of the inner core of Earth, it has been proposed that compositional convection could take place due to the precipitation of magnesium-bearing minerals  \cite{o2016powering}. The key difference between these two buoyancy sources lies in the contrast of the thermal diffusivity, $\kappa_{T}$, and the compositional diffusivity, $\kappa_{C}$.
The co-existence of the thermal and compositional buoyancy gives rise to the so-called double-diffusive convection (DDC), which is thought to occur in Earth's core today, with thermal convection (TC) and compositional convection (CC) contributing approximately 20\% and 80\% of the geodynamo's power, respectively \cite{lister1995strength}. DDC is also considered to be relevant to Mercury's core \cite{manglik2010dynamo,takahashi2019mercury}, where significant stable layers may exist to account for the weak magnetic field of Mercury \cite{Johnson2018}. Therefore, studying dynamos driven by DDC is crucial for understanding core dynamics and magnetic field generation in terrestrial planets.

DDC has been extensively studied in the context of oceanography \cite{radko2013double}, where temperature and salinity serve as distinct sources of buoyancy for seawater. In oceanography, most of studies on DDC adopt local Cartesian box models \cite<e.g.,>[]{simeonov2008double,Ouillon_Edel_Garaud_Meiburg_2020,Li_Yang_2024}. 
However, global spherical models are more relevant to the dynamics of planetary cores. In this context, early studies on DDC have focused on the hydrodynamic onset of convection \cite{busse2002low,trumper2012numerical,net2012numerical,silva2019onset}. These studies have shown that the properties of the critical onset mode, such as its drift frequency or azimuthal wave number, strongly depends on buoyancy sources. In the highly supercritical regime, direct numerical simulations have shown that the flow morphology in DDC can be rather different from that of single buoyancy driven convection \cite{breuer2010thermochemically,trumper2012numerical}. 
This is not unexpected in the hydrodynamic regime because thermal convection alone highly depends on the Prandtl number, the ratio of momentum diffusivity to thermal diffusivity, as previous studies have shown \cite{Aurnou_Bertin_Grannan_Horn_Vogt_2018,Abbate2023,Fan_Wang_Lin_2024}. The question then arises as to whether the distinct convective flow morphologies of TC, CC, and DDC are preserved in the presence of magnetic fields, particularly in the strong-field regime relevant to Earth's core. 

Several studies have investigated self-consistent dynamos driven by DDC via direct numerical simulations. \citeA{takahashi2014double} found that the morphology of the dynamo-generated magnetic field is related to the ratio of thermal and compositional buoyancy power injection. \citeA{mather2021regimes} reported that the onset dynamos in the TC-dominated and CC-dominated regimes exhibit distinct magnetic morphologies. Dynamos driven by DDC involving top stable layers were also numerically studied in the context of Mercury's dynamo \cite{manglik2010dynamo,takahashi2019mercury}. More recently, \citeA{tassin2021geomagnetic} conducted a systematic parameter survey of DDC dynamo simulations and shown that Earth-like magnetic fields can be generated for any partition of the thermal and compositional power input. However, they also found that the transition from the dipole-dominated to multipolar dynamos highly depends on the ratio of  the power input between thermal and compositional contributions. These studies suggest that it would be desirable to consider DDC dynamos instead of the widely adopted co-density model in dynamo simulations \cite{wicht2019advances}. The co-density (COD) model proposed by \citeA{braginsky1995equations} assumes the effective diffusivity values for $\kappa_{T}$ and $\kappa_{C}$ are equal, which has the advantage of saving computational costs by reducing the number of control parameters and governing equations. The COD model has proven to be successful in reproducing Earth-like magnetic fields and secular variations \cite{christensen2010conditions,aubert2013bottom,schaeffer2017turbulent} and approaching realistic dynamic regimes \cite{aubert2023}. In the strong-field dynamo regime, the magnetic field may mitigate the influence of different buoyancy sources due to the dominant role of the Lorentz force \cite{Dormy_2016,schwaiger2021,yuanlonghui-F}. However, given the large contrast between $\kappa_{T}$ and $\kappa_{C}$ in planetary cores, the validity of the co-density assumption remains to be tested \cite{wicht2019advances}. 

In this study, we conduct a series of numerical simulations of double-diffusive convection and self-consistent dynamos driven by DDC in a rotating spherical shell. Additionally, we compare the DDC models with the corresponding co-density models. The primary motivation of this work is to test the validity of the co-density assumption in the dynamic regime relevant to planetary cores. 
Therefore, we focus exclusively on the top-heavy DDC regime, in which both the thermal and compositional gradients are destabilizing. Broadly speaking, DDC also includes two other configurations: the so-called fingering convection, where the compositional gradient is unstable but the thermal gradient is stable \cite<e.g.,>{monville2019rotating,guervilly2022fingering}, and the so-called semi-convection, where the thermal gradient is unstable but the compositional gradient is stable \cite<e.g.,>{net2012numerical,monville2019rotating}. For a comprehensive discussion of these two configurations, we refer the reader to the review article by \citeA{Garaud2018}.

Our DDC models have two distinct features compared to previous studies on the top-heavy convection. First, our DDC models adopt a large value of the Lewis number, $Le=100$, which represents the ratio of the thermal diffusivity ($\kappa_{T}$) to the compositional diffusivity ($\kappa_{C}$). The Lewis number is a crucial control parameter in a DDC system because the key difference of two buoyancy sources lies in the contrast of diffusivities \cite{tassin2021geomagnetic}.
While the diffusivity parameters in planetary cores remain uncertain \cite<e.g.,>{loper1981study,braginsky1995equations}, it is widely accepted that thermal diffusivity is much larger than compositional diffusivity, with the Lewis number potentially reaching $10^2\sim 10^{4}$ \cite{BOUFFARD2017552,wicht2019advances}. However, DDC simulations at large $Le$ are computationally demanding because of the scale differences between the thermal and compositional fields \cite{BOUFFARD2017552}. Most previous DDC simulations typically adopted $Le=10$ due to numerical limitations. Second, for the DDC dynamo models, we focus on the strong-field regime, where the magnetic energy significantly exceeds the kinetic energy, in order to investigate the magnetic effects on DDC. Our simulations reveal that convective flow morphologies differ significantly between the single-diffusive models and the DDC models in the hydrodynamic regime as previous studies have shown. However, these differences are largely mitigated by the presence of magnetic fields in the dynamo models. The magnetic fields and secular variations generated by the DDC models and the co-density models at the core-mantle boundary are remarkably similar. Our numerical experiments provide further justifications for the use of the co-density assumption in planetary dynamo simulations.   

The remainder of this paper is organized as follows: Section \ref{sec:models} introduces the numerical models and methods, while Section \ref{sec:results} presents the numerical results. Finally, the paper concludes with a summary and discussion in Section \ref{sec:conclusions}.

\section{Numerical Models} \label{sec:models}
\subsection{Governing Equations}
 We consider Boussinesq convection of homogeneous fluid in a spherical shell of inner radius $r_i$ and outer radius $r_o$ that uniformly rotates at $\boldsymbol \Omega=\Omega \boldsymbol{\hat z}$. DDC is driven by fixed temperature difference $\Delta T=T_{i}-T_{o}$ and fixed composition difference $\Delta C=C_{i}-C_{o}$ between the inner and outer boundaries, under the gravity $\boldsymbol{g}=-g_0 \boldsymbol{r}/r_o$, where $g_0$ is the gravity at the outer boundary. The magnetic field is matched to a potential field outside the fluid domain (the inner-core and mantle are both modelled as electrically
 insulating). We adopt the shell thickness $D = r_{o} - r_{i}$ as the length scale,  $\Omega^{-1}$ as the time scale, $\Delta T$ as the temperature scale, $\sqrt{\rho \mu \eta \Omega}$ as the magnetic scale, where $\mu$ is magnetic permeability and $\eta$ is magnetic diffusivity. The dimensionless governing equations can be expressed as
 \begin{linenomath*}
 \begin{equation}
  \frac{\partial \boldsymbol u}{\partial t}+ \boldsymbol u \cdot \boldsymbol \nabla\boldsymbol u+2 \hat{\boldsymbol z} \times \boldsymbol u = -\boldsymbol \nabla P+E \boldsymbol \nabla^{2} \boldsymbol u+ ({Ra_{T}^*}T + {Ra_{C}^*}C \boldsymbol) \frac{\boldsymbol r}{r_o}+ \frac {E}{Pm}[(\boldsymbol \nabla\times \boldsymbol B)\times \boldsymbol B],
  \label{eq21}
 \end{equation}
 \end{linenomath*}
 \begin{linenomath*}
 
 \begin{equation}
  \frac{\partial T}{\partial t}+\boldsymbol u \cdot \boldsymbol \nabla T =\frac{E}{Pr} \boldsymbol \nabla^{2} T,
  \label{eq22}
 \end{equation}
 
 \end{linenomath*}
 \begin{linenomath*}
 \begin{equation}
  \frac{\partial C}{\partial t}+\boldsymbol u \cdot \boldsymbol \nabla C =\frac{E}{Sc} \boldsymbol \nabla^{2} C,
  \label{eq23}
 \end{equation}
 \end{linenomath*}
 \begin{linenomath*}
 \begin{equation}
  \frac{\partial \boldsymbol B}{\partial t} =\boldsymbol \nabla\times (\boldsymbol u\times \boldsymbol B)+ \frac{E}{Pm}\boldsymbol \nabla^{2}\boldsymbol B,
  \label{eq24}
 \end{equation}
 \end{linenomath*}
 \begin{linenomath*}
 \begin{equation}
  \boldsymbol \nabla \cdot \boldsymbol u=0,
  \label{eq25}
 \end{equation}
 \end{linenomath*}
 \begin{linenomath*}
 \begin{equation}
  \boldsymbol \nabla \cdot \boldsymbol B=0,
  \label{eq26}
 \end{equation}
 \end{linenomath*}
 where $\boldsymbol{u}$ is the velocity, $P$ is the reduced pressure, $T$ is the temperature $C$ is the composition and $\boldsymbol{B}$ is magnetic. The system is defined by six dimensionless control parameters, the Ekman number $E$, the modified thermal Rayleigh number $Ra_{T}^*$, the modified compositional Rayleigh number $Ra_{C}^*$, the Prandtl number $Pr$, the Schmidt number $Sc$ and magnetic Prandtl number $Pm$: 
 \begin{linenomath*}
 \begin{equation}
  E=\frac{\nu}{\Omega D^2},\quad Ra_{T}^*=\frac{\alpha g_{0}\Delta T}{\Omega^2 D}, \quad Ra_{C}^*=\frac{\beta  g_{0}\Delta C}{\Omega^2 D}, \quad  Pr=\frac{\nu}{\kappa_{T}}, \quad Sc=\frac{\nu}{\kappa_{C}}, \quad Pm=\frac{\nu}{\eta},
  \label{eq27}
 \end{equation}
 \end{linenomath*}
 where $\nu$ is the kinematic viscosity, $\alpha$ is the thermal expansion coefficient and $\beta$ is the compositional expansion coefficient. The rotationally modified thermal or compositional Rayleigh number $Ra_{T/C}^*$ is often used in the rotating convection, which is also the squared convective Rossby number \cite{Gilman01011977}. Note that the Lewis number $Le$ mentioned in Section \ref{sec:intro} is the ratio of the Schmidt number to the Prandtl number,
 \begin{linenomath*}
 \begin{equation}
  Le=\frac{\kappa_{T}}{\kappa_{C}}=\frac {Sc}{Pr}.
  \label{eq28}
 \end{equation}
 \end{linenomath*}
 In the COD model, the Rayleigh number $Ra^*$ takes the same form as $Ra_{T/C}^*$. The buoyancy term in equation \ref{eq21} is consolidated into a single term, while equations \ref{eq22}-\ref{eq23} are merged into a unified equation.
 
 In this study, we fix the radius ratio of the shell $\eta=r_i/r_o=0.35$. Both boundaries are impermeable, electrically insulating, no-slip and maintained at constant temperatures and compositions. While a zero-flux condition for composition at the CMB would be more realistic for planetary interiors \cite{tassin2021geomagnetic,guervilly2022fingering}, we choose to fix both temperature and composition at the CMB to facilitate comparison with the co-density model. The initial velocity field is set to zero, while the initial temperature field, composition field, and magnetic field initialized with random perturbations. 
 
 \subsection{Numerical method}
 We use the open-source code XSHELLS (\url{https://www.bitbucket.org/nschaeff/xshells/}) to solve the governing equations (\ref{eq21}-\ref{eq26}) subjected to the boundary conditions. The fluid is assumed to be incompressible, and the velocity $\boldsymbol u$ and magnetic $\boldsymbol B$ can be decomposed into toroidal and poloidal components:
 \begin{linenomath*}
 \begin{equation}
 \boldsymbol u=\boldsymbol \nabla \times(T\boldsymbol r)+\boldsymbol \nabla \times \boldsymbol \nabla \times (P\boldsymbol r),
 \label{eq29}
 \end{equation}
 \end{linenomath*}
 \begin{linenomath*}
 \begin{equation}
 \boldsymbol B=\boldsymbol \nabla \times(\mathcal{T} \boldsymbol r)+\boldsymbol \nabla \times \boldsymbol \nabla \times(\mathcal{P} \boldsymbol r).
 \label{eq210}
 \end{equation}
 \end{linenomath*}
 The toroidal $T$ and $\mathcal{T}$, poloidal $P$ and $\mathcal{P}$ scalar fields, as well as the temperature field $T$ and compositional field $C$, are expanded using spherical harmonic expansion on spherical surfaces. XSHELLS employs a second-order finite differences method in the radial direction, along with a pseudo-spectral spherical harmonic expansion. The spectral expansion is truncated up to spherical harmonics of degree $l_{max}$ and $Nr$ denotes the number of radial grid points. The time-stepping scheme is second order, and treats the diffusive terms implicitly, while the nonlinear and Coriolis terms are handled explicitly.  To enhance computational efficiency, the code utilizes the SHTns library for fast spherical harmonic transformations \cite{schaeffer2013efficient}. XSHELLS has been benchmarked against other geodynamo codes \cite{marti2014full, matsui2016performance} and has previously been applied to study DDC in full-sphere geometry \cite{monville2019rotating}.

 \subsection{Diagnostics}
 We analysis several diagnostics properties to quantify the effect of various control parameters on heat, composition and momentum transports, as well as magnetic field strength and morphology. The Nusselt (Sherwood) number $Nu$ $(Sh)$ denotes heat (composition) transport, the ratio of the total heat (composition) flux to the conduction heat (composition) flux. In spherical shells, the conductive temperature (compositional) profile $T_{c}$ $(C_{c})$ is the solution of
 \begin{linenomath*}
 \begin{equation}
 \frac{\mathrm{d} T_{c}}{\mathrm{~d} r} = -\frac{r_{i}r_{o}}{r^2}, \quad T_{c}\left(r_{i}\right) = 1, \quad T_{c}\left(r_{o}\right) = 0,
 \label{eq211}
 \end{equation}
 \end{linenomath*}
 \begin{linenomath*}
 \begin{equation}
 \frac{\mathrm{d} C_{c}}{\mathrm{~d} r} = -\frac{r_{i}r_{o}}{r^2}, \quad C_{c}\left(r_{i}\right) = 1, \quad C_{c}\left(r_{o}\right) = 0.
 \label{eq212}
 \end{equation}
 \end{linenomath*}
 Following \citeA{gastine2015turbulent}, it yields
 \begin{linenomath*}
 \begin{equation}
 T_{c}(r)=C_{c}(r)=\frac{\eta}{(1-\eta)^2}\frac{1}{r}-\frac{\eta}{1-\eta},
 \label{eq213}
 \end{equation}
 \end{linenomath*}
 where $\eta=r_{i}/r_{o}$ is the radius ratio. The notation $\vartheta$ and $\varpi$ are introduced to define the time and horizontally averaged radial dimensionless temperature and compositional profile, respectively
 \begin{linenomath*}
 \begin{equation}
 \vartheta(r)=\overline{\langle T\rangle_s}, \quad \varpi(r)=\overline{\langle C\rangle_s},
 \label{eq214}
 \end{equation}
 \end{linenomath*}
 where ${\overline{\dots}}$ correspond to temporal averaging, $\langle \dots\rangle_s$ to an average over a spherical surface. Then the Nusselt and Sherwood number are
 \begin{linenomath*}
 \begin{equation}
 Nu=-\eta\frac{\mathrm{d}\vartheta}{\mathrm{d}r}(r=r_{i})=-\frac{1}{\eta}\frac{\mathrm{d}\vartheta}{\mathrm{d}r}(r=r_{o}), \quad Sh=-\eta\frac{\mathrm{d}\varpi}{\mathrm{d}r}(r=r_{i})=-\frac{1}{\eta}\frac{\mathrm{d}\varpi}{\mathrm{d}r}(r=r_{o}).
 \label{eq215}
 \end{equation}
 \end{linenomath*}
 The magnetic $Eb$ and kinetic $Eu$ are given by
 \begin{linenomath*}
 \begin{equation}
  Eb=\frac{1}{2Pm/E} \int_{v} \boldsymbol B^2 \mathrm{d}V, \quad Eu=\frac{1}{2} \int_{v} \boldsymbol u^2 \mathrm{d}V,
  \label{eq216}
 \end{equation}
 \end{linenomath*}
 where $V$ is spherical shell of volume. The kinetic $Eu$ can be decomposed into zonal and non-zonal parts:
 \begin{linenomath*}
 \begin{equation}
  Eu_{zon}=\frac{1}{2} \int_{v}{(u_{\phi}^{0})}^2 \mathrm{d}V, \quad Eu_{non}=Eu-Eu_{zon},
  \label{eq217}
 \end{equation}
 \end{linenomath*}
 where $u_{\phi}^0$ is the axisymmetric ($m = 0$) component of the azimuthal velocity $u_{\phi}$. We define the Rossby and Reynolds number as
 \begin{linenomath*}
 \begin{equation}
  Ro=\frac{U_{rms}}{\Omega D}, \quad Re=Ro/E,
  \label{eq218}
 \end{equation}
 \end{linenomath*}
 where $U_{rms}$ is the dimensional root-mean-square (r.m.s.) velocity. Based on the non-dimensionalisation we used, the Reynolds number can be determined through kinetic energy as
 \begin{linenomath*}
 \begin{equation}
  Re=\frac{1}{E}\sqrt{\frac{2Eu}{V}}.
  \label{eq219}
 \end{equation}
 \end{linenomath*}
 Accordingly, we have the zonal Reynolds and non-zonal Reynolds number 
 \begin{linenomath*}
 \begin{equation}
  Re_{zon}=\frac{1}{E}\sqrt{\frac{2Eu_{zon}}{V}}, \quad Re_{non}=\frac{1}{E}\sqrt{\frac{2Eu_{non}}{V}}.
  \label{eq220}
 \end{equation}
 \end{linenomath*}
 The flow field length scale $L_u$ is determined from the time-averaged kinetic energy spectrum following \citeA{christensen2006scaling}
\begin{equation}
L_u^{-1} = \overline{\left(\frac{D \sum\limits_{l = 1}^{l_{\max }} \sum\limits_{m = 0}^{l} l \mathcal{E}_{l}^{m}(t)}{\pi \sum\limits_{l = 1}^{l_{\max }} \sum\limits_{m = 0}^{l} \mathcal{E}_{l}^{m}(t)}\right)},
\label{eq38}
\end{equation}
 where $\mathcal{E}_{l}^{m}$ is the dimensionless kinetic energy at a spherical harmonic degree $l$ and order $m$. The magnetic field length scale $L_b$ is also determined following the same definition as that of the flow field length scale. The magnetic Reynolds number can be determined through Reynolds number and magnetic Prandtl number as
 \begin{linenomath*}
 \begin{equation}
  Rm=RePm.
  \label{eq221}
 \end{equation}
 \end{linenomath*}
 The thermal and compositional buoyancy power $P_T$ and $P_C$ defined by
 \begin{linenomath*}
 \begin{equation}
  P_{T}(t) =V\left\langle{Ra_T^*} T(\mathbf{r}, t) \frac{r}{r_{o}} u_{r}(\mathbf{r}, t)\right\rangle_V, \quad P_{C}(t) =V\left\langle{Ra_C^*} C(\mathbf{r}, t) \frac{r}{r_{o}} u_{r}(\mathbf{r}, t)\right\rangle_V,
  \label{eq222}
 \end{equation}
 \end{linenomath*}
 where $\langle \dots\rangle_V$ correspond to an average over a spherical shells. We define the Elsasser number as 
 \begin{linenomath*}
 \begin{equation}
  \Lambda =\frac{B_{rms}^2}{\rho \mu \eta \Omega},
  \label{eq223}
 \end{equation}
 \end{linenomath*}
 where $B_{rms}$ is the dimensional root-mean-square (r.m.s.) magnetic. Finally, for the $f_{dip}$, we following \citeA{christensen2006scaling}, the fraction of axial dipole in the observable spectrum (up to $l=13$).

 \section{Results}\label{sec:results}
 \subsection{Hydrodynamics simulations}\label{sec:Hydrodynamics simulations}
 We begin by conducting hydrodynamic double-diffusive convection (HD-DDC) simulations, with the primary aim of analyzing the interaction between TC and CC. To begin with, we perform separate single-diffusive convection simulations for pure TC and pure CC to establish baseline reference cases. By comparing the results of the subsequent HD-DDC simulations with these baselines, we can clearly identify the changes in TC and CC that occur under HD-DDC. Regarding the selection of the Rayleigh numbers for pure TC and pure CC, we ensure that the resulting buoyancy powers are within a comparable range and that the convective intensity remains insufficient to break the rotational constraint. To mitigate potential randomness in the conclusions drawn from a single numerical experimental group and to enhance the reliability of our results, we conduct two groups of numerical experiments. 
 The primary difference between these two groups lies in the buoyancy powers of TC and CC, with the second group exhibiting greater buoyancy power. During the numerical experiments, we fixed $E=3 \times 10^{-5}$, $Pr=0.1$, $Sc=10$, while the buoyancy power is adjusted by varying the parameters $Ra_T^{*}$ and $Ra_C^{*}$.

 \begin{figure}
 \noindent\includegraphics[width=\textwidth]{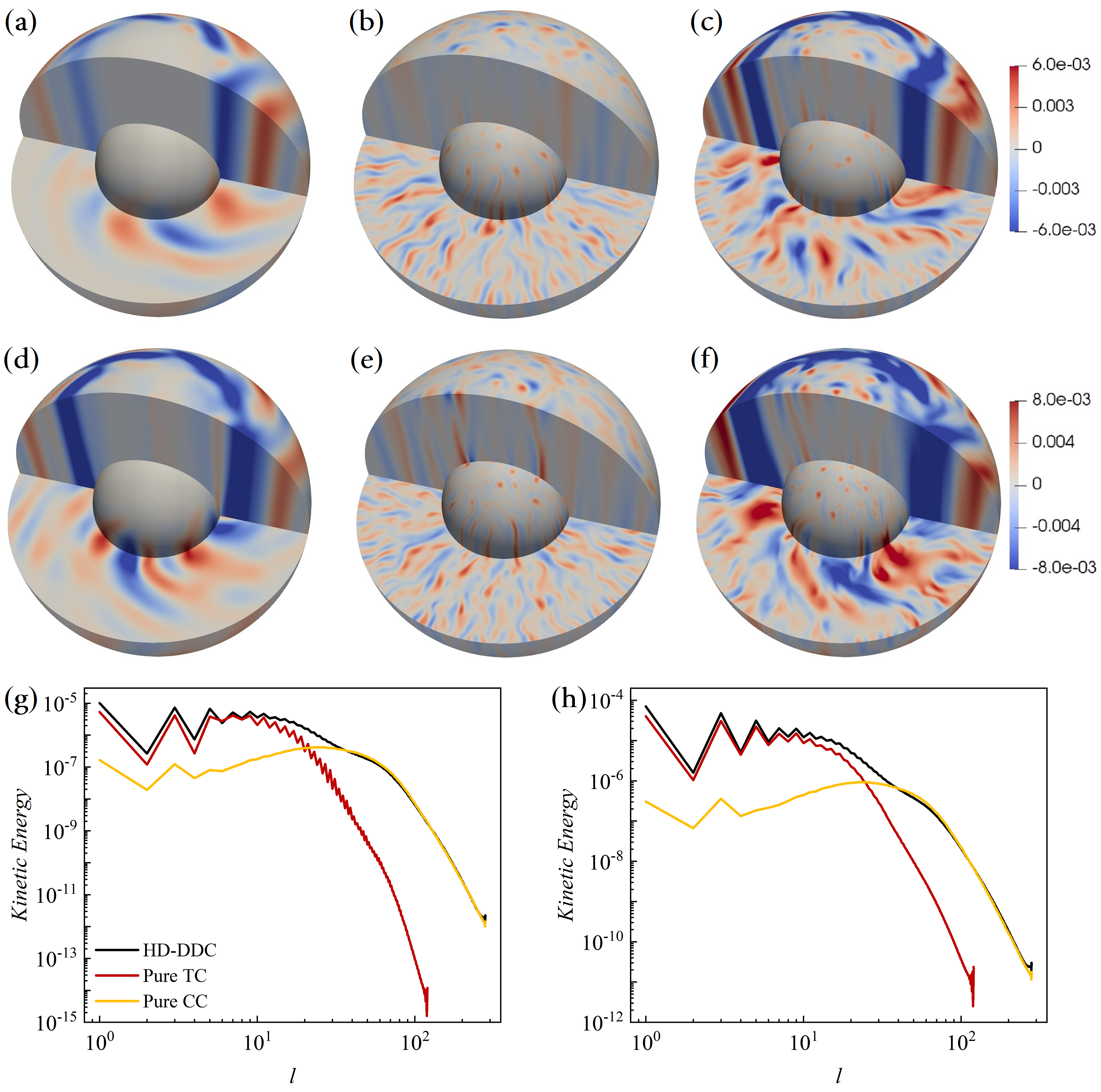}
 \caption{3-D flow fields renderings and energy spectra of hydrodynamic simulations in Table \ref{table1}. The inner (outer) surface corresponds to a spherical surface of radius $r_i+0.1D$ $(r_o-0.1D)$. The equatorial plane and inner core surface show the radial velocity $u_r$. The meridional plane and outer core surface show the azimuthal velocity $u_\phi$. The color bar represents Rossby number $Ro$. Panels (a-c, g) correspond to  Group 1; panels (d-f, h) correspond to Group 2. The Left column shows the pure thermal convection cases; the middle column shows the pure compositional convection cases, and the right column shows the double-diffusive convection cases.}
 \label{pure_to_DDC}
 \end{figure}
 
\begin{table}
 \caption{Parameters and diagnostics of our hydrodynamics simulations. Pure TC and Pure CC represent the hydrodynamic single-diffusive thermal convection and single-diffusive compositional convection, respectively. HD-DDC represents the hydrodynamic double-diffusive convection. The arrows symbolize the variations in transport and buoyancy power in HD-DDC relative to pure TC or pure CC.}
 \centering
 \begin{tabular}{l cccccc}
 \hline
  &&Group 1&&&Group 2&\\
  ~~Model&Pure TC&Pure CC&HD-DDC&~~~Pure TC&Pure CC&HD-DDC\\
 \hline
   ~~${Ra_T^*}$ & 0.018 & 0 & 0.018 & ~~~0.03 & 0 & 0.03\\
   ~~${Ra_C^*}$ & 0 & 0.009 & 0.009 &~~~ 0 & 0.015 & 0.015 \\
   ~~~$Re$ & 84.2 & 51.8 & 118 & ~~~184 & 77.9 & 238\\
   ~$Re_{zon}$ & 50.9 & 8.21 & 68.0 & ~~~126 & 12.4 & 158\\
   ~$Re_{non}$ & 67.1 & 51.1 & 96.4 &~~~ 134 & 76.9 & 178\\
   ~~~$L_{u}$ & 0.35 & 0.09 & 0.22 & ~~~0.39 & 0.09 & 0.29\\
   ~$Nu-1$ & 0.04 & 0 & $0.07 \uparrow$ & ~~~0.11 & 0 & $0.15 \uparrow $\\
   ~$Sh-1$ & 0 & 9.41 & $8.66 \downarrow $ & ~~~0 & 13.8 & $12.4 \downarrow $\\
   ${P_T}(\times 10^{-6})$ & 0.95 & 0 & $1.73\uparrow$ & ~~~4.58 & 0 & $7.28\uparrow $\\
   ${P_C}(\times 10^{-6})$ & 0 & 1.81 & $1.65\downarrow$ & ~~~0 & 4.41 & $3.95 \downarrow $\\
 \hline
 \label{table1}
 \end{tabular}
 \end{table}
 
 As shown in Figure \ref{pure_to_DDC}, significant differences in flow morphology between pure TC and pure CC are clearly observed in both groups of numerical experiments. Pure TC exhibits a larger length scale in the radial flow along the equatorial plane and stronger azimuthal flow in the meridional plane. This behavior is consistent with previous studies on rotating convection at low Prandtl numbers \cite{aubert2001systematic,Fan_Wang_Lin_2024}. In contrast, pure CC is predominantly characterized by small-scale plumes in the equatorial plane, a typical feature of convection patterns at lower compositional diffusivity \cite{Li2021,Abbate2024}, with weaker azimuthal flow in the meridional plane. When TC and CC blend to form HD-DDC, the resulting flow morphology combines characteristics of both pure TC and pure CC. Near the outer boundary of the equatorial plane, small-scale flows typical of CC is observed, while large-scale flows characteristic of TC dominates the interior. In the meridional plane, the azimuthal flow intensity of HD-DDC is stronger than that of pure TC. Figures \ref{pure_to_DDC}(g) and (h) show the energy spectra corresponding to the two groups of numerical experiments. It can be seen that the spectrum of the HD-DDC model is essentially a superposition of those from the pure TC and pure CC cases, supporting the qualitative observations from the flow field morphology. However, further quantitative analysis is needed to determine whether the HD-DDC model is merely a simple linear combination of the pure TC and pure CC contributions. 

 An analysis of Table \ref{table1} reveals that, in both groups of numerical experiments, when the buoyancy powers are similar, the $Re_{non}$ of pure TC and pure CC are relatively close. However, the $Re_{zon}$ of pure TC is significantly higher, and its flow exhibits a larger length scale, as clearly seen in Figure \ref{pure_to_DDC}. To further explore whether pure TC and pure CC assume a simple superposition relationship within the HD-DDC model, it is necessary to consider factors such as heat and composition transport, as well as the buoyancy powers of TC and CC. In both groups of numerical experiments, TC is notably enhanced in the HD-DDC model, with heat transport and thermal buoyancy power nearly doubling. This is primarily attributed to the significant enhancement of the heat transfer process by the radial flow of CC. In contrast, CC has not been enhanced in the same manner as TC. In fact, both compositional transport and compositional buoyancy power are somewhat reduced. We believe that this is due to the strong zonal flow of TC, which impedes the radial compositional transport.

Through quantitative analysis, we find that in the HD-DDC model, TC and CC do not follow a simple superposition pattern. Specifically, CC significantly enhances TC, while TC exerts a slight inhibitory effect on CC. This difference in their interactions  is predominantly rooted in the differences in their flow morphology. It is important to note that above-mentioned conclusions were drawn under the condition of $Le=100$. When the value of $Le$ is relatively small, the differences in the flow morphology between TC and CC will correspondingly decrease. In such cases, TC  may no longer inhibit CC. Conversely, when $Le$ is relatively large, the differences in the flow morphology of TC and CC convection will be further magnified. As a result, the promoting and suppressing effects between them are likely to become even more prominent.

 \subsection{Dynamo simulations}
 In the previous section, we analyzed the interactions between TC and CC in the HD-DDC model. 
 In this section, we investigate the further influence of a strong magnetic field on the DDC model by comparing it with the HD-DDC cases discussed earlier. The choice to study the strong magnetic field regime is based on previous studies \cite{gillet2010fast,finlay2011flow}, which suggests that the ratio of magnetic energy to kinetic energy ($Em/Ek$) in Earth's dynamo lies between $10^3$ and $10^4$. 
There are two main approaches to achieving a strong-field dynamo regime in numerical simulations: one involves reducing the magnetic Ekman number \cite{yadav2016approaching, schwaiger2019force}, and the other involves increasing the magnetic Prandtl number $Pm$ \cite{Dormy_2016}. Here, we follow the latter approach by adopting a relatively large $Pm=10$, while using the same control parameters as in the HD-DDC model from the previous section. For dynamo simulations, we introduce the Roberts number $q$, which is the ratio of thermal (or compositional) diffusivity to the magnetic diffusivity: 
 
 \begin{linenomath*}
 \begin{equation}
  q =\frac{\kappa_{T/C}}{\eta}=\frac{Pm}{Pr} \quad or =\frac{Pm}{Sc}.
  \label{eq31}
\end{equation}
 \end{linenomath*}
 Due to the large $Pm$ employed in our simulations, the Roberts number is 
$q=100$ for TC and $q=1$ for CC.

 \begin{table}[t]
 \caption{Parameters and diagnostics of our hydrodynamics and dynamos simulations. HD-DDC and MHD-DDC represent the hydrodynamic double-diffusive convection and double-diffusive convection dynamo, respectively.}
 \centering
 \begin{tabular}{l cccc}
 \hline
  & \multicolumn{2}{l}{~~~~~~~~~~~~Group 1} & \multicolumn{2}{l}{~~~~~~~~~~~~Group 2} \\
   ~~Model&HD-DDC&MHD-DDC&~~HD-DDC&MHD-DDC\\
 \hline
   ~~~${Ra_T^*}$ & 0.018 & 0.018 & ~~~0.03 & 0.03\\
   ~~~${Ra_C^*}$ & 0.009& 0.009 & ~~~0.015 & 0.015\\
   ~~~${Pm}$ & 0& 10 & ~~~0 & 10\\
   ~$Em/Ek$ & 0 & 30.1 & ~~~0 & 31.8\\
   ~~~$Re$ & 118 & 43.7 & ~~~238 & 114\\
   ~~$Re_{zon}$ & 68.0 & 11.6 & ~~~158 & 38.0\\
   ~~~$L_{u}$ & 0.22 & 0.16 & ~~~0.29 & 0.23\\
   ~$Nu-1$ & 0.07 & 0.03$\downarrow$ & ~~~0.15& 0.28$\uparrow$\\
   ~$Sh-1$ & 8.66 & 8.15$\downarrow$ & ~~~12.4& 13.2$\uparrow$\\
   ${P_T}(\times 10^{-6})$& 1.73 & 0.88$\downarrow$ & ~~~7.28 & 15.6$\uparrow $\\
   ${P_C}(\times 10^{-6})$& 1.65& 1.56$\downarrow$ &~~~3.95 & 4.24$\uparrow $\\
 \hline
 \label{table2}
 \end{tabular}
 \end{table}

\begin{figure}[t]
 \noindent\includegraphics[width=\textwidth]{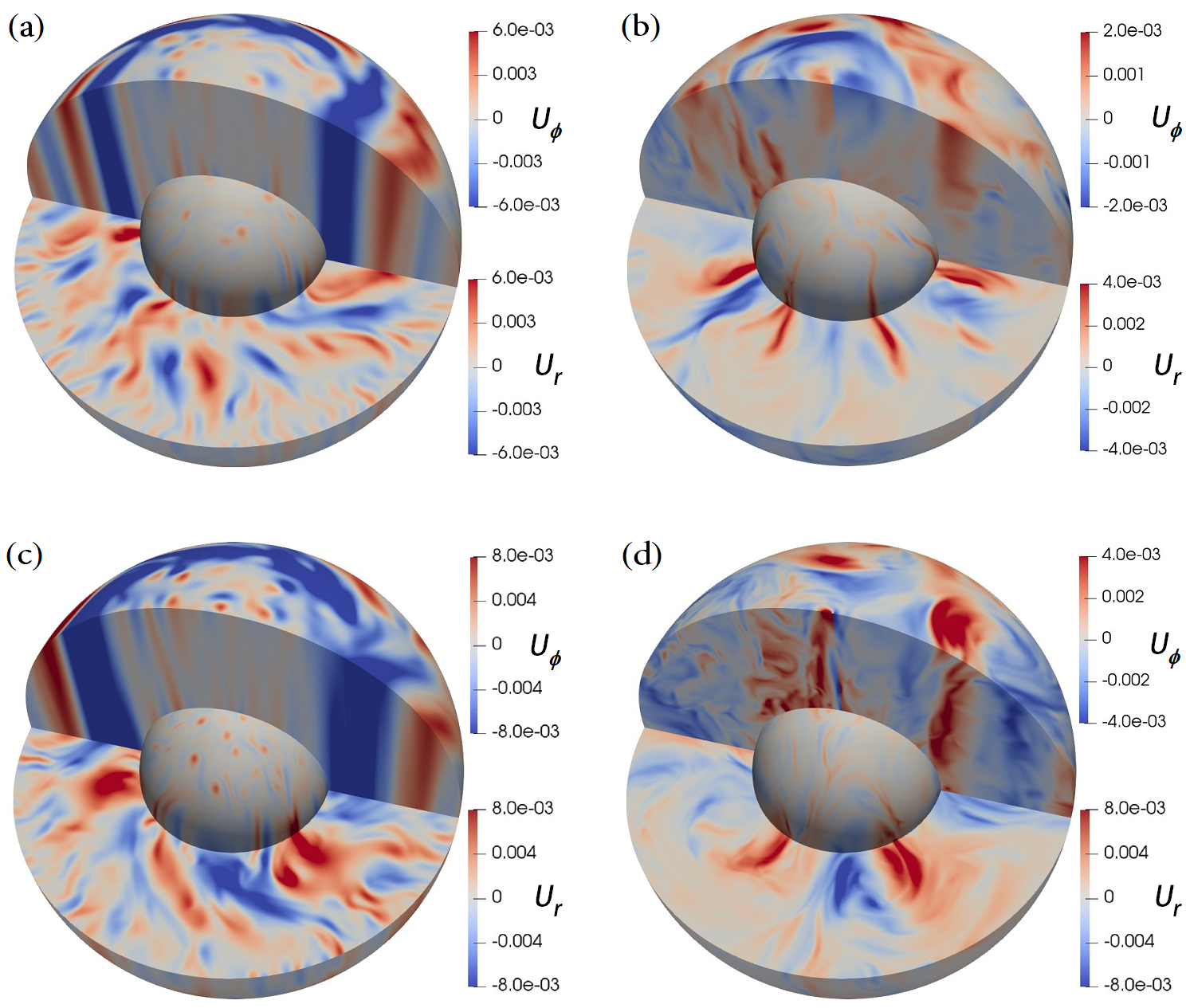}
 \caption{Renderings of 3-D flow fields from the hydrodynamic and dynamo simulations listed in Table~\ref{table2}. This figure follows the format of Figure~\ref{pure_to_DDC}, but panels (a) and (c) show hydrodynamic double-diffusive models, while panels (b) and (d) correspond to double-diffusive convection dynamo models.The first row corresponds to group 1 and second row corresponds to group 2.}
 \label{DDCvsMHD}
 \end{figure}

As shown in Table \ref{table2}, the Reynolds numbers $Re$ in both groups of simulations decrease significantly under the influence of a strong magnetic field, indicating suppressed flow intensity. Figure \ref{DDCvsMHD}(b) and (d) further illustrates that the convective columns are disrupted in both cases, suggesting that the magnetic field exerts a similar structural impact on the flow in both cases. However, the effect of the strong magnetic field on buoyancy power differs markedly between the two groups of simulations. To investigate this, we conducted single-diffusive dynamo tests for the pure TC and pure CC cases corresponding to groups 1 and 2 in Table~\ref{table1}. In group 1, the pure TC is insufficient to generate a magnetic field, indicating that the dynamo process in the MHD-DDC model (group 1, Table~\ref{table2}) is primarily driven by CC. Consequently, the dominant Roberts number in this system is $q=1$ (i.e., $Pm/Sc=1$). In group 2, both pure TC and pure CC are capable of sustaining dynamo process. However, we can see from Table \ref{table2} that thermal buoyancy power dominates over the compositional buoyancy power in the MHD-DDC case (group 2, Table~\ref{table2}), implying that the dominant Roberts number $q=100$ in this case. Based on these results, we suggest that the differential feedback of the magnetic field on buoyancy power may be influenced by the difference in the dominant Roberts number between the two groups \cite{SIMITEV_BUSSE_2005,Dormy2024}.

The changes in buoyancy power induced by the strong magnetic field lead to a shift in the dominant mode of the MHD-DDC models. In group 1, the dynamo is dominated by compositional buoyancy power, whereas in group 2, thermal buoyancy power becomes dominant. However, due to the regulating effect of the Lorentz force under strong magnetic fields, the characteristic flow features observed in the HD-DDC models disappear in both cases (as shown in Figure \ref{DDCvsMHD} (b) and (d)). The equatorial plane displays the typical radially elongated flow structures commonly associated with strong-field convection, regardless of the dominant buoyancy source. This suggests that under strong magnetic field conditions, variations in the buoyancy source may not have a significant impact on the overall dynamo process.

 \subsection{Comparison with the co-density model}
In the previous section, we analyzed the effects of a strong magnetic field on DDC and identified the key role played by the Roberts number $q$ in the dynamo process. In this section, we will compare the MHD-DDC model constructed in the previous section with the co-density dynamo (MHD-COD) model. The aim is to explore the differences in the dynamo processes driven by different buoyancy forces, thereby enabling us to assess the suitability of the COD model. In the COD model, we assume no difference between thermal and compositional diffusivity, setting $Pr=Sc=1$. Additionally, two important conditions must be satisfied for the MHD-COD model: firstly, the model should be in a strong magnetic field regime, which is consistent with the actual magnetic field conditions on Earth. Secondly, due to the different Prandtl numbers in the COD and DDC models, the critical Rayleigh numbers required to initiate convection differ significantly. As a result, it is not feasible to compare the two models by fixing the Rayleigh number. Instead, we fix the magnetic Prandtl number and select cases with similar $Rm$ for comparison. The $Rm$ is defined by the ratio of the magnetic freezing term to the magnetic dissipation term in Equation (equation \ref{eq24}). 

 \begin{table}[t]
 \caption{Parameters and diagnostics of our dynamos simulations. MHD-COD and MHD-DDC represent the co-density dynamo model and double-diffusive convection dynamo, respectively}
 \centering
 \begin{tabular}{l cccc}
 \hline
  & \multicolumn{2}{l}{~~~~~~~~~~~~~Group 1} & \multicolumn{2}{l}{~~~~~~~~~~~~~~~Group 2} \\
   ~~~~~Model&MHD-COD&MHD-DDC&~~MHD-COD&MHD-DDC\\
 \hline
   ~~~~~~${Ra^*}$ &0.012 & / & ~~0.03 & /\\ 
   ~~~~~~${Ra_T^*}$ & / & 0.018 & ~~/ & 0.03\\  
   ~~~~~~${Ra_C^*}$ & / & 0.009 & ~~/ & 0.015\\
   ~~~~~~$Pr$ & 1 & 0.1 & ~~1 & 0.1\\
   ~~~~~~$Sc$ & 1 & 10 & ~~1 & 10\\
   ~~~~~~$Pm$ & 10 & 10 & ~~10 & 10\\
   ~~~~$Em/Ek$ & 74.7 & 30.1 & ~~20.5 & 31.8\\
   ~~~~~~$Re$ & 43.8 & 43.7 & ~~113 & 114\\
   ~~~~~$Re_{zon}$ & 13.7 & 11.6 & ~~30.3 & 38.0\\
   ~~~~~~$Rm$ & 438 & 437 & ~~1125 & 1137\\
   ~~~~~~~$L_{u}$ & 0.24 ($\pm$ 0.02) & 0.16 ($\pm$ 0.01)& ~~0.16 ($\pm$ 0.01) & 0.23 ($\pm$ 0.02)\\
   ~~~~~~~$L_{b}$ & 0.26 ($\pm$ 0.01)& 0.18 ($\pm$ 0.01)& ~~0.14 ($\pm$ 0.01) & 0.20 ($\pm$ 0.01)\\
   ~·~~~~~$\Lambda$ & 43.8 & 17.2 & ~~78.0 & 123\\
   ~~~~~$\Lambda_{CMB}$ & 3.95 & 1.54 & ~~7.78 & 10.1\\
   ~~~~~~$f_{dip}$ & 0.64 ($\pm$ 0.02)& 0.68 ($\pm$ 0.02)& ~~0.45 ($\pm$ 0.03)& 0.52 ($\pm$ 0.02)\\
   ~~~~$\frac{P_T}{P_T+P_C}$&/&0.36&~~/&0.79 \\
   ~~~~~~~$N_{r}$ & 200& 240 & ~~280 & 300\\
   ~~~~~~~$l_{max}$ & 200& 240 & ~~280 & 300\\
   million.core.hours & $\simeq 0.04$& $\simeq 0.1$ & ~~$\simeq 0.18$ & $\simeq 0.5$\\
 \hline
 \label{table3}
 \end{tabular}
 \end{table}

 \begin{figure}
 \noindent\includegraphics[width=\textwidth]{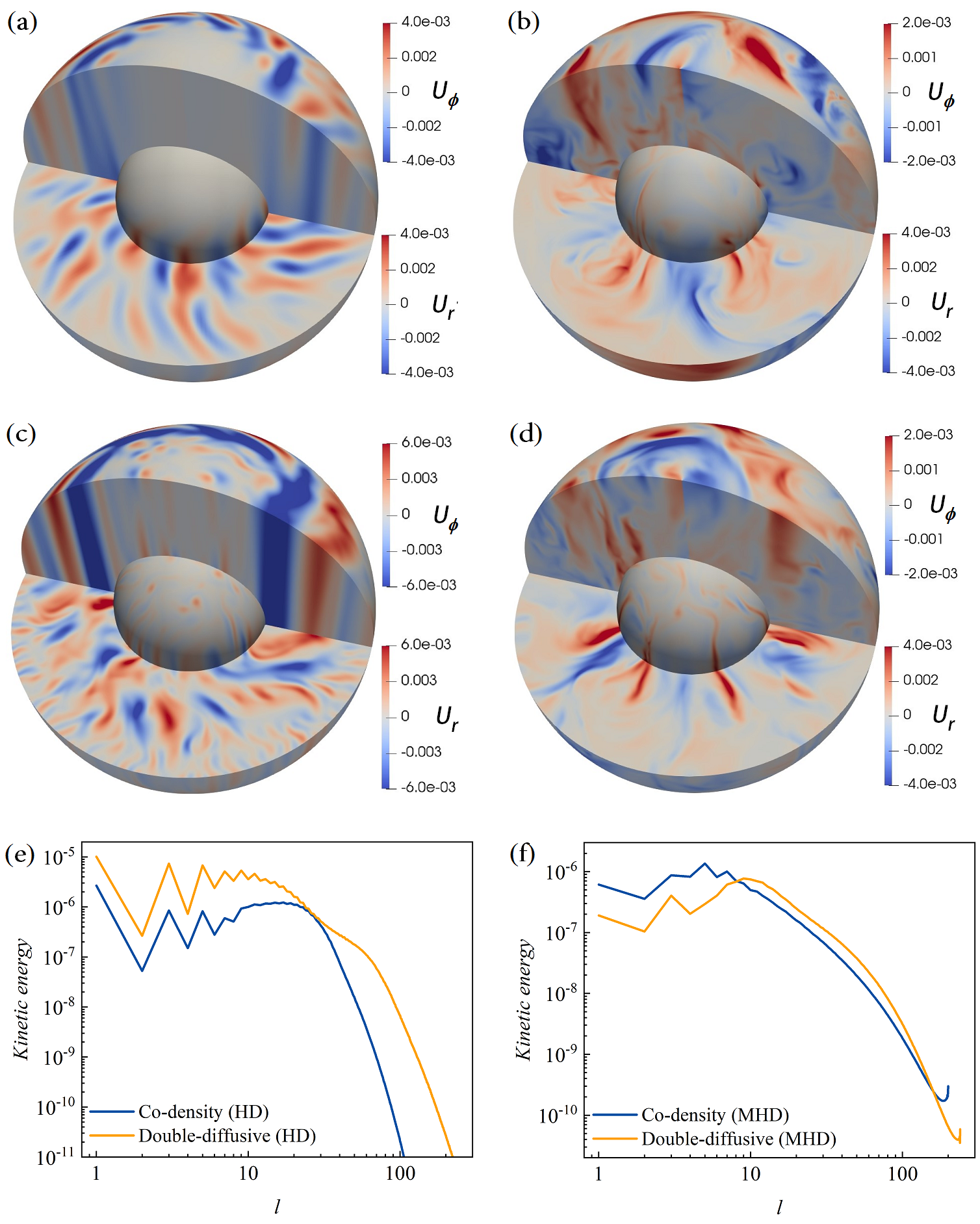}
 \caption{Comparison of flow fields between the co-density model and the double-diffusive
model. This figure follows the format of Figure~\ref{pure_to_DDC}. Panels (a) and (c) show the hydrodynamic results of the co-density and double-diffusive models, respectively, while panels (b) and (d) display the corresponding dynamo results. Panels (e) and (f) present the time-averaged kinetic energy spectra for the hydrodynamic and dynamo cases, respectively.  The dynamo cases correspond to the Group 1 in Table \ref{table3}.}
 \label{Rm500Ek}
 \end{figure}

 \begin{figure}
 \noindent\includegraphics[width=\textwidth]{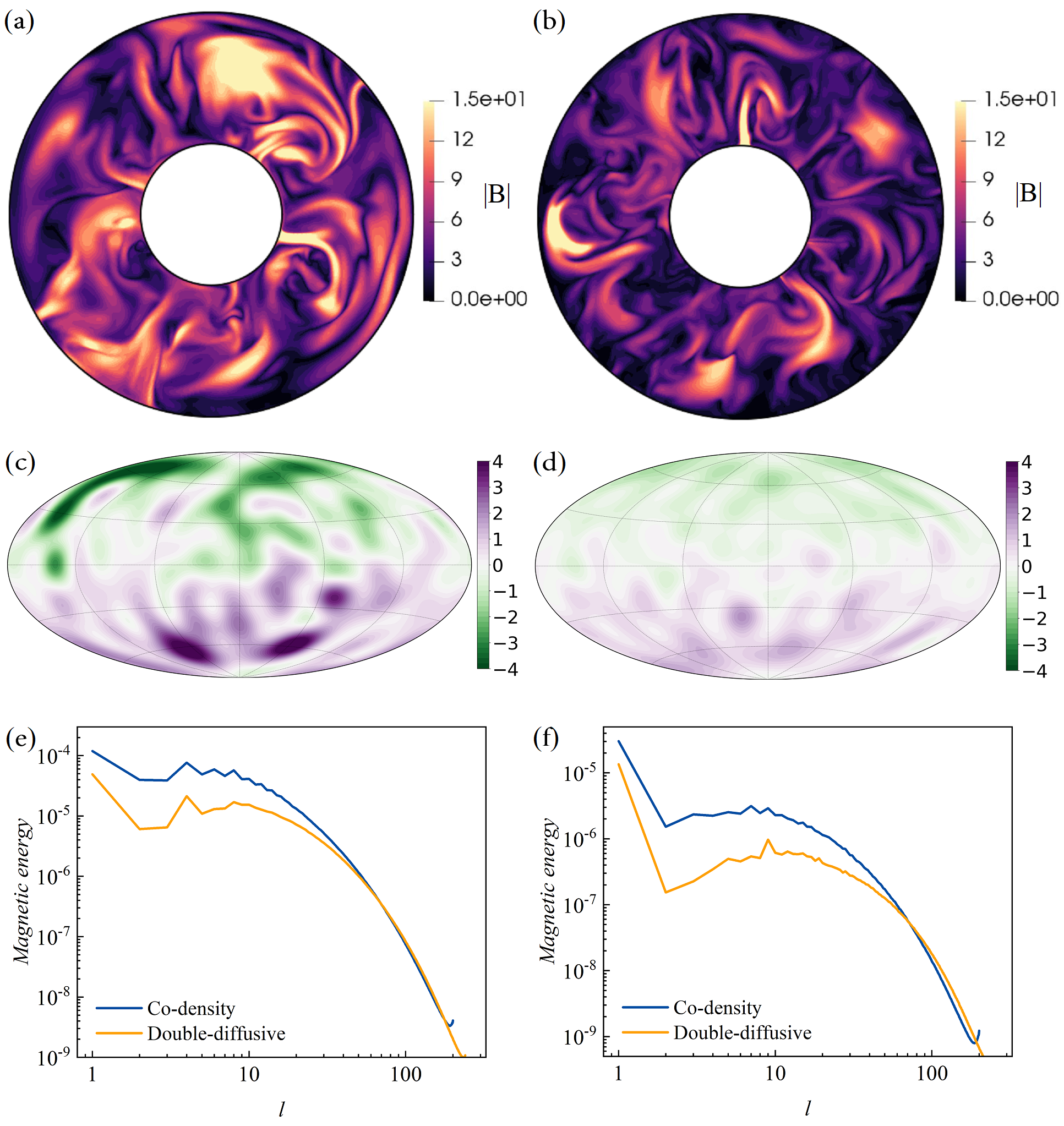}
 \caption{Comparison of magnetic fields between the co-density model and the double-diffusive model. Panels (a) and (b) show the magnetic field intensity in the equatorial plane for the co-density model and the double-diffusive model, respectively. Panels (c) and (d) present the radial magnetic field at the core-mantle boundary, truncated at $l=13$, in aitoff projection for the co-density model and the double-diffusive model, respectively. Panel (e) shows the time-averaged magnetic energy spectrum, and Panel (f) displays the time-averaged magnetic energy spectrum at the core-mantle boundary. Note that the magnetic field is in (non-dimensional) Elsasser units (scaled by $\sqrt{\rho \mu \eta \Omega}$). These figures correspond to Group 1}.
 \label{Rm500Em}
 \end{figure}
 
 \subsubsection{Dynamo model with $Rm \approx 500$}
 In the group 1 of comparisons presented in Table \ref{table3}, the $Rm$ of both models are approximately 500. However, the MHD-COD model exhibits a higher magnetic-to-kinetic energy ratio and a larger Elsasser number $\Lambda$, indicating a stronger Lorentz force acting on the system. In the MHD-DDC model, since the compositional buoyancy dominates, the flow length scale is smaller than that in the MHD-COD model, and correspondingly, the magnetic field also exhibits a smaller length scale. Additionally, the $f_{dip}$ values of both models exceed 0.5, indicating that both are in a dipole-dominated regime.

 However, the diagnostic parameters presented above are insufficient to definitively determine whether the two models are similar. Therefore, we consider further verification from the perspective of the energy spectrum. Figures \ref{Rm500Ek}(a) and (c) show the hydrodynamic results of the COD and DDC models, respectively, while Figures \ref{Rm500Ek}(b) and (d) display the results under strong magnetic fields. It is evident that under hydrodynamic conditions, the flow morphologies of the two models differ significantly. This is also reflected in the kinetic energy spectra shown in Figure \ref{Rm500Ek}(e), where the energy distributions across different scales are notably different. Under the influence of a strong magnetic field, differences in flow structures and spectra (Figure \ref{Rm500Ek}(f)) still persist. However, compared to the hydrodynamic case, these differences are significantly reduced due to the dominant role of the Lorentz force in regulating the flow, and both models exhibit radially elongated flow structures. 
 
 We then proceed to analyze the magnetic fields generated by the two models. As shown in Figures \ref{Rm500Em}(a) and (b), the MHD-COD model exhibits a stronger magnetic field at the equatorial plane than the MHD-DDC model. At the core-mantle boundary, the radial magnetic field strength also differs between the two (Figure \ref{Rm500Em}(c) and (d)), though the magnetic field morphology appears quite similar. Comparison of the magnetic energy spectra, including those at the CMB, reveals that although the absolute magnitudes of magnetic energy differ between the models, their spectral trends are highly consistent. This indicates a similar distribution of magnetic energy across scales.  Therefore, we conclude that the magnetic fields generated by the MHD-COD model and the MHD-DDC model dominated by compositional buoyancy are qualitatively similar, despite minor differences in their magnetic field length scales. The case in which thermal buoyancy dominates will be examined in the next section.

  \begin{figure}
 \noindent\includegraphics[width=\textwidth]{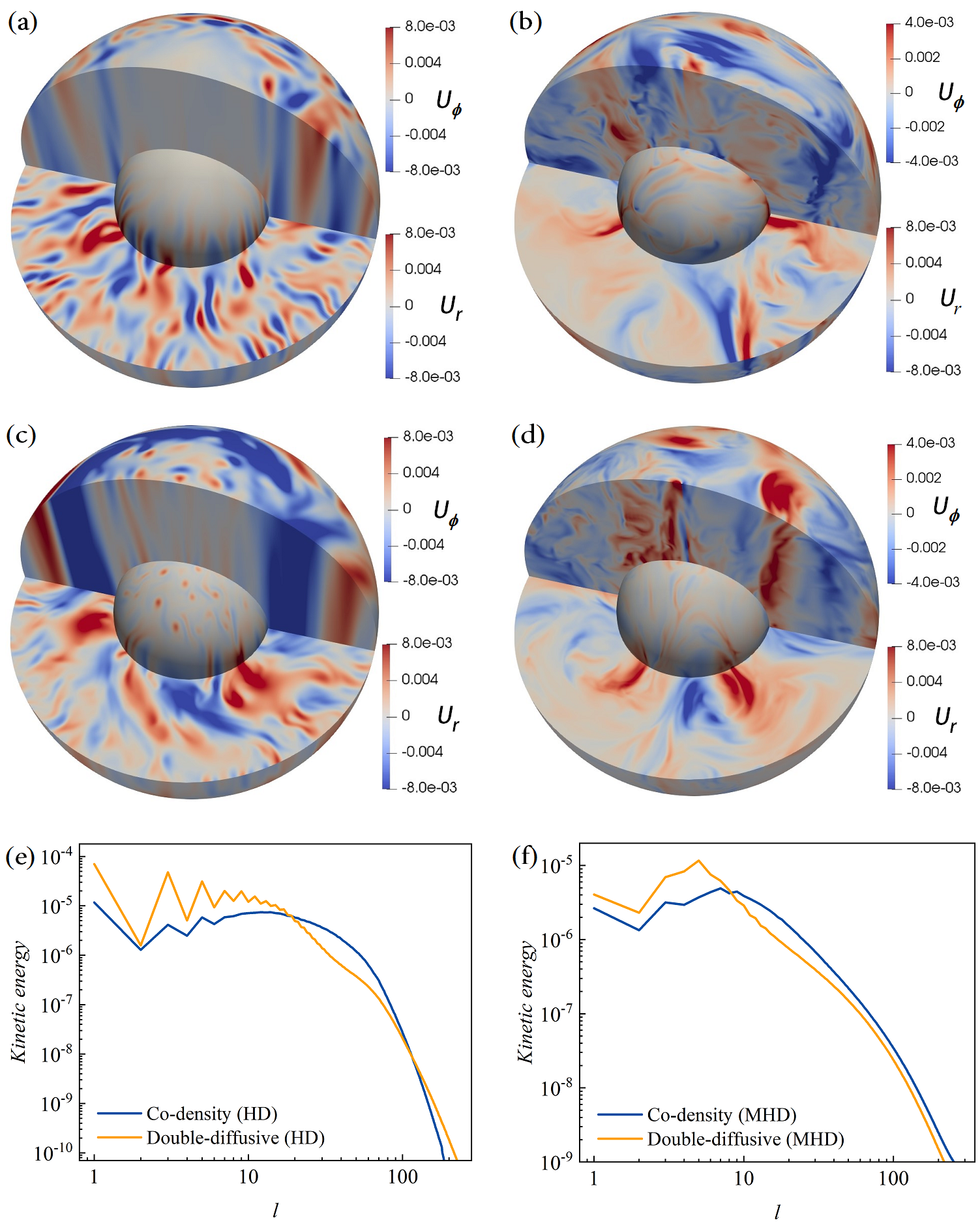}
 \caption{Comparison of flow fields between the co-density model and the double-diffusive
model. This figure follows the format of Figure~\ref{Rm500Ek}, but for cases with a magnetic Reynolds number of approximately 1000.  The dynamo cases correspond to the Group 2 in Table \ref{table3}.}
 \label{Rm1000Ek}
 \end{figure}

\subsubsection{Dynamo model with $Rm \approx 1000$}
In this section, we compare the MHD-DDC model driven primarily by TC, as described earlier, with the MHD-COD model. 
From the group 2 of data in Table \ref{table3}, it is evident that both models have magnetic Reynolds numbers around 1000. In terms of the flow and magnetic fields, the MHD-DDC model dominated by thermal buoyancy exhibits larger length scales. This is consistent with the large-scale flow structures typically associated with thermal convection under hydrodynamic conditions. Furthermore, the $\Lambda$ of the MHD-DDC is relatively higher, indicating that the system is subjected to a stronger Lorentz force. Although both models are in a strong magnetic field, due to the relatively high magnetic Reynolds numbers, their $d_{dip}$ are both in the dipolar-multipolar transition range.

\begin{figure}[t]
 \noindent\includegraphics[width=\textwidth]{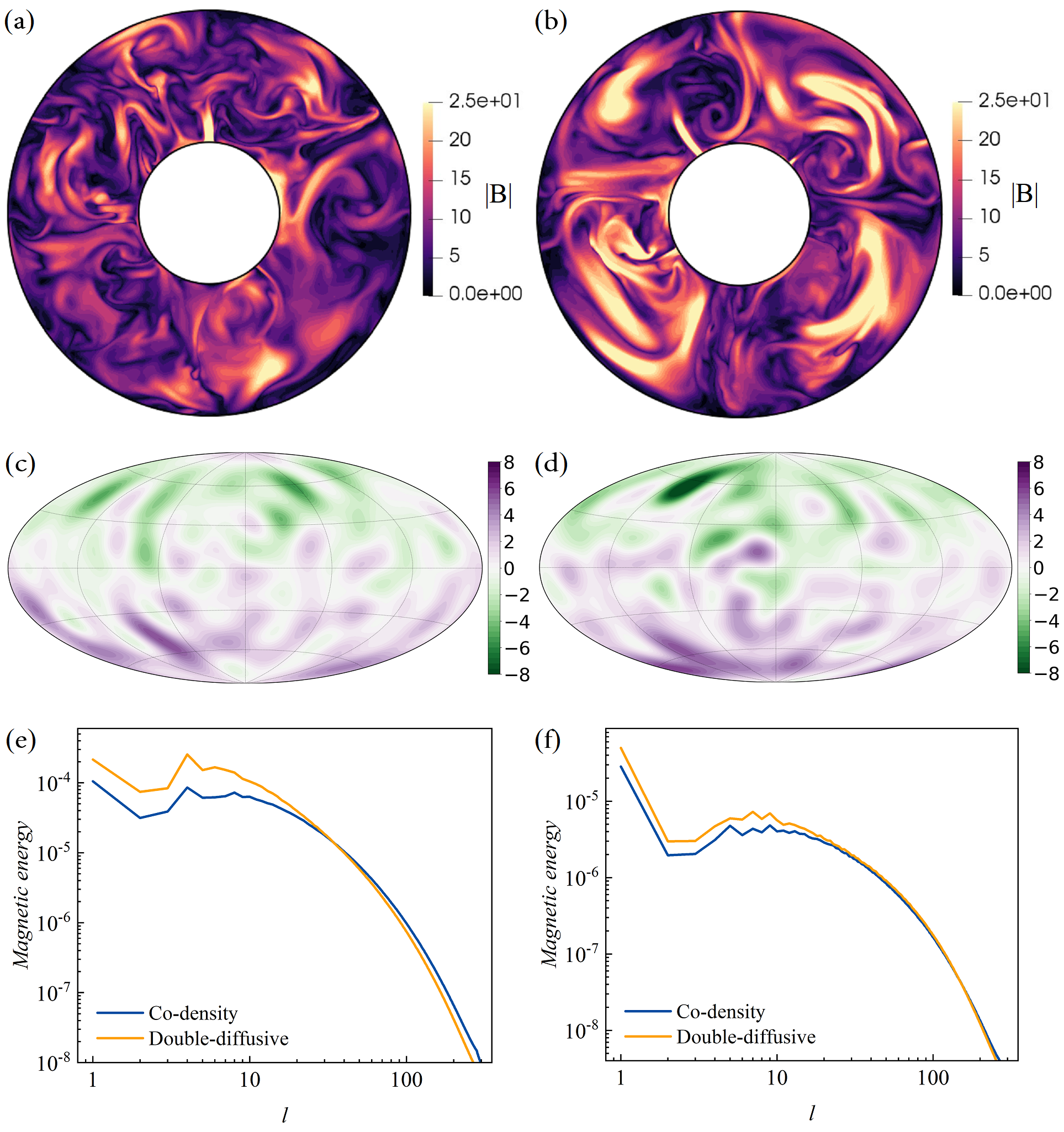}
 \caption{Comparison of magnetic fields between the co-density model and the double-diffusive model. This figure follows the format of Figure~\ref{Rm500Em}, but for cases with a magnetic Reynolds number of approximately 1000.  These figures correspond to Group 2.}
 \label{Rm1000Em}
 \end{figure}

Subsequently, we continue with further verification from the perspective of the energy spectrum. As shown in Figures \ref{Rm1000Ek} (a) and (c), the flow fields of the two models still exhibit clear differences under hydrodynamic conditions. These differences are also evident in the kinetic energy spectra shown in Figure \ref{Rm1000Ek} (e), where the energy distributions across different scales differ significantly. Similar to the results from the group 1, the flow morphologies and spectral trends of the two models remain distinct; however, under the influence of a strong magnetic field (as shown in figure \ref{Rm1000Ek} (b) and (d)), these differences are substantially reduced. Next, we analyze the generated magnetic fields. As shown in Figures \ref{Rm1000Em} (a) and (b), the two models exhibit some differences in both the magnetic field strength at the equatorial plane and the magnetic length scale. However, with respect to the radial magnetic field at the CMB (see Figures \ref{Rm1000Em} (c) and (d)), the two models display a high degree of similarity. Finally, from the global magnetic energy spectra and those at the CMB, we observe that both models exhibit similar spectral trends, with the MHD-DDC model showing slightly higher energy at large scales. Therefore, when thermal buoyancy dominates, the magnetic fields generated by the MHD-DDC and MHD-COD models remain qualitatively similar, although differences in magnetic field length scales persist.

 \begin{figure}
 \noindent\includegraphics[width=\textwidth]{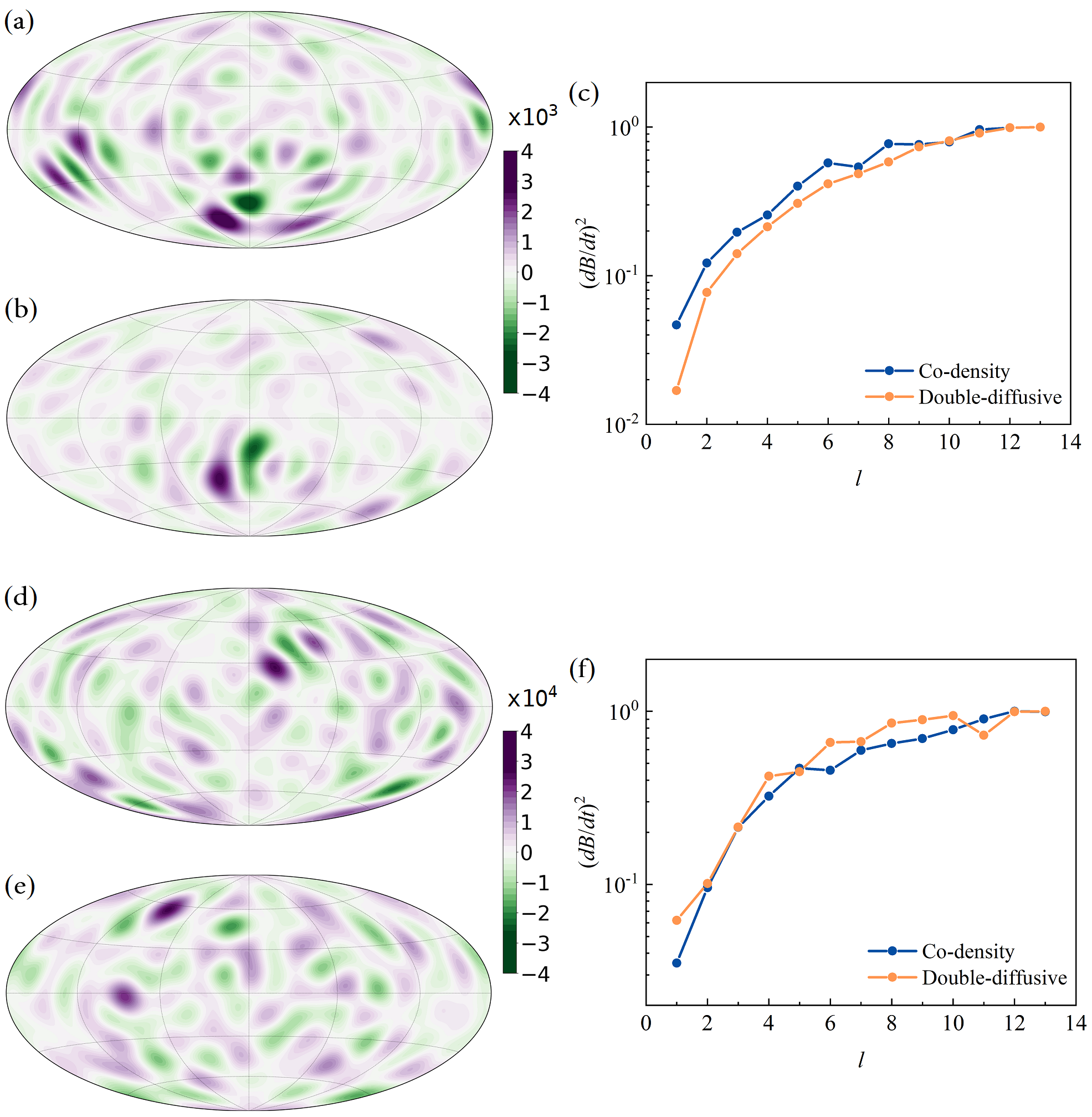}
 \caption{Aitoff projections of the secular variation at the core–mantle boundary, truncated at spherical harmonic degree $l = 13$. Panels (a) and (b) correspond to the co-density and double-diffusive model with $Rm \approx 500$, respectively. Panel (c) shows the time-averaged secular variation spectrum at the core-mantle boundary with $Rm \approx 500$. Panels (d) and (e) correspond to co-density and double-diffusive model, with $Rm \approx 1000$, respectively.  Panel (f) shows  the time-averaged secular variation spectrum at the core-mantle boundary with $Rm \approx 1000$. Note that the secular variation spectrum has been normalized by with the maximum spherical harmonic degree.}
 \label{SV}
 \end{figure}
 
 \subsubsection{Secular variations}
 Based on the previous analysis, we conclude that under strong magnetic fields and comparable $Rm$, dynamo processes driven by different buoyancy types do not exhibit significant differences. In this section, we further analyze the magnetic fields generated by the MHD-DDC and MHD-COD models. Our focus is on the secular variation of the magnetic field, defined as the first time derivative of the field. We aim to explore potential differences in dynamo processes driven by different buoyancy from the perspective of magnetic field observations. As shown in Figures \ref{SV} (a) and (b), when the $Rm$ is approximately 500, the magnetic field variations in the CC-dominated MHD-DDC model, and the MHD-COD model are quite similar, both exhibiting significant north-south asymmetry. Figure \ref{SV} (c) displays the time-averaged energy spectrum of the secular variations at the CMB, where it is evident that the trends in the energy spectra of both models are similar. This further confirms that the two models exhibit comparable secular variations. When the $Rm$ is approximately 1000, the TC-dominated MHD-DDC model, and the MHD-COD model not only show similar snapshots of secular variations, but also display comparable trends in their energy spectra. This consistency suggests that, whether the MHD-DDC is primarily driven by TC or CC, the secular variations do not exhibit substantial differences when compared to the MHD-COD model. Based on the above-mentioned comparisons, we can reasonably infer that magnetic field observations alone may not be sufficient to clearly identify the dominant convection mode in Earth's liquid outer core.

\section{Conclusions}\label{sec:conclusions}
In this study, we conducted a series of numerical simulations of both hydrodynamics and dynamos driven by double-diffusive convection to explore the interactions between thermal convection and compositional convection. In our hydrodynamic simulations, we found that the convective flow morphologies differ significantly among pure thermal convection, pure compositional convection, and double-diffusive convection with comparable buoyancy power injection. The hydrodynamic double-diffusive convection simulations reveal that the strong zonal flows driven by thermal convection tend to suppress radial compositional transport, whereas the small-scale radial flows driven by compositional convection enhance heat transfer. These interactions underscore the complexity of convective systems in astrophysical and geophysical contexts at large Lewis numbers.

By comparing the hydrodynamic and dynamo simulations of double-diffusive convection, we found that the differential impact of the strong magnetic field on buoyancy power in the two groups is associated with the dominant buoyancy source, and is therefore likely related to the Roberts number. Meanwhile, under the regulating effect of the strong magnetic field, differences in flow morphology are largely suppressed by the dominant Lorentz force, regardless of which buoyancy source is dominant.

Finally, we conducted a comparative analysis between the double-diffusive convection dynamo model and the co-density dynamo model. Under strong magnetic field conditions and comparable magnetic Reynolds numbers, we found that the magnetic fields generated by the double-diffusive convection model are qualitatively similar to those produced by the co-density model, despite some differences in magnetic field strength and length scales. The difference in buoyancy sources between the two models does not significantly affect the distribution of magnetic energy across different scales. This suggests that the co-density model remains applicable for studying geodynamo processes. 
Moreover, the similarity in secular variation implies that it is difficult to infer the underlying buoyancy source of a planetary magnetic field solely from observations at or above the planetary surface. 

While these findings provide insights into the influence of buoyancy source and applicability of the co-density mode, we acknowledge that our conclusions are based on a limited number of simulations. A more comprehensive comparison across a broader range of parameters may help assess the extent to which these findings can be generalized.

\section*{Data Availability Statement}
Numerical simulations made use of the open-source code XSHELLS \cite{schaeffer2013efficient}. Input parameters and diagnostic output parameters of numerical simulations are given in Tables \ref{table1}-\ref{table3}. For more comprehensive and detailed data, an Excel file is available on Zenodo \cite{dataset}.

\acknowledgments
This study was supported by the National Key R\&D Program of China (2022YFF0503200) and the National Natural Science Foundation of China (42174215,12250012). YL is also supported by the Pearl River Program (2019QN01X189).
Numerical calculations were performed on the Taiyi cluster supported by the Center for Computational Science and Engineering of Southern University of Science and Technology.

%
%

%
%
%
%
%

\end{document}